\documentclass[twocolumn]{article}

\newcommand{\beq}{\begin{equation}}
\newcommand{\eeq}{\end{equation}}
\newcommand{\ket}[1]{\left| {#1} \right>}
\newcommand{\bra}[1]{\left< {#1} \right|}

\newcommand{\micron}{\;\mu{\rm m}}

\usepackage{graphics}

\begin{document}

\title{Quantum computing with trapped ions, atoms and light}
\author{A. M. Steane and D. M. Lucas \\
{\small Centre for Quantum Computation,}\\
{\small Department of Physics, University of Oxford,}\\
{\small Clarendon Laboratory, Parks Road, Oxford OX1 3PU, England.}}
\date{\today}
\maketitle

\begin{abstract}
We first consider the basic requirements for a quantum computer, arguing for the
attractiveness of nuclear spins as information-bearing entities, and
light for the coupling which allows quantum gates. We then survey the strengths of
and immediate prospects for quantum information processing in ion traps.
We discuss decoherence and gate rates in ion traps, comparing methods
based on the vibrational motion with a method based on exchange of photons
in cavity QED. We then sketch the main features of a quantum computer 
designed to allow an algorithm needing $10^6$ Toffoli gates on
100 logical qubits. We find that around 200 ion traps linked by
optical fibres and high-finesse cavities
could perform such an algorithm in a week to a month, using components
at or near current levels of technology.
\end{abstract}

\section{Introduction}

This paper will discuss various issues in the physics of
ion traps and quantum information. 
Our purpose is to address some quite general questions
about quantum information physics and ion (or atom) traps, with the aim
of identifying useful directions for theoretical and experimental
research in the near future and the longer term.

We begin in section \ref{s_ideal} by considering the major requirements
of a quantum computer without assuming any particular technology from
the outset. Rather, we try to identify physical phenomena which appear
to be intrinsically well-suited to quantum computing, using arguments
based as much as possible on fundamental physical principles. We find that
there is a natural link with methods in quantum optics, such as ion and
atom trapping and cooling. In section \ref{s_iontrap} we focus our attention
on currently achieved experimental results, considering the particular
strengths of ion trap methods. Section \ref{s_short} briefly surveys
the main experiments in quantum information physics which are feasible
in the near future using ion traps. These include several fundamental
quantum information effects which have not yet been observed in any
area of physics. In section \ref{s_gate} we examine
how far ion trap methods can go: we estimate the major limitations to
the gate rate, at given precision, for two different methods to implement
the gates. These are the coupling via the vibrational degree of freedom
which has been used up till now, and coupling via cavity quantum
electrodynamics (CQED) methods. In section \ref{s_design} we
sketch a design for a moderately large quantum computer: one which
could perform $10^6$ Toffoli gates on 100 logical qubits. This
computer, based on atomic physics and CQED concepts, is conceivable
using current technology, and the quantum optics methods it is based on
are probably necessary in any case for quantum communication links. It
illustrates the power of these methods, and underlines the interest of
further experiments, and theoretical studies, in this area.

\section{An ideal physical system for quantum computing}  \label{s_ideal}

We would like to consider the question, what
might be the ideal system for a future quantum computer?
Here, we do not intend to restrict attention to any one branch of physics (or other science).
Rather, we want to know what system we might choose, if we are guided only by basic
physical principles and the properties of systems which, in some useful sense, Nature provides.

We would like our ideal quantum computer to have the highest quality memory, logic
gates, and read-out. We note that the
read-out gives automatically the ability to prepare simple initial states such as
product states. By ``highest quality'' we mean primarily reliable operation,
but if the system or the gate can also be small or fast, then so much the better.

For a quantum memory, we want a system of qubits
which does not interact with anything else,
while for logic gates we want a coupling between qubits which is fast, and not
coupled to anything other than the bits. These two demands are almost, but not quite,
contradictory. They imply that a quantum computer should be composed of entities $q$ which
are, in their passive state when no gates are switched on, almost completely
isolated, and yet which can be coupled rapidly. This means they must have a strong
coupling to something, $\cal G$. The conflicting demands are met if $\cal G$ has the property
that it can be made to be wholly absent when it is not wanted, and introduced rapidly
when it is wanted. Furthermore we would like $\cal G$ to interact only with the
entities $q$ and with nothing else.

It turns out that Nature does provide a physical entity $q$ which meets the
contradictory demands of memory and logic gates better than we might have imagined
possible. This entity is the nuclear spin. 
The advantages of nuclear spins for quantum computing are already well recognised.
A spin has a smaller coupling 
to its environment than any degree of freedom based on charge or the motion of
particles; a nuclear spin has a particularly small magnetic moment,
and this tiny magnet comes ready-packaged in an electron cloud with
highly useful properties for logic gates. The atomic electrons provide the means
to take hold of the atom and place it where we wish, and they also provide a ready-made very
strong and very stable magnetic field on the nucleus. This results in the hyperfine
splitting. The stability of this splitting for isolated 
atoms or ions is well documented, and is in fact used
to provide our standards of time and frequency.

The existence of hyperfine structure makes it possible to couple
to the nuclear spin via the electronic state. This provides the handle whereby
logic gates can be achieved. The next question is, what is the best way to grasp
this handle?

The existing proposals which are based on the nuclear spin
and/or hyperfine splitting
differ in the way the coupling $\cal G$ is brought about.
These proposals include bulk 
nuclear magnetic resonance (NMR) \cite{93:Lloyd,96:Cory,97:Gershenfeld},
ion trap \cite{95:Cirac} and other atomic-physics-based
methods \cite{95:Pellizzari,99:Jaksch}, and a
proposal for future solid state devices based on nuclear spins of dopant
atoms implanted in a semiconductor \cite{98:Kane}.

In bulk liquid-state NMR, the electronic `handle' is almost entirely ignored,
and the method relies instead on the tiny direct
spin-spin interaction between neighbouring nuclei in a molecule. This permits
logic gates but not a direct measurement of the spin state. The fact that all
neighbouring qubits are permanently coupled in such methods has both advantages
and disadvantages. The other methods
all use the electronic `handle'. Ion trap experiments up till now have used
a light-induced coupling between the electronic state and the vibrational motion
(phonon) of relatively heavy charged particles (ions) in a trap in
vacuum \cite{95:Cirac,95:Monroe,98:Turchette,00:Sackett,99:Roos}. There are proposals
in which light alone is used to couple the electronic state
of one atom and another \cite{95:Pellizzari,98:Cirac}, and a realisation of this
concept (in an experiment not based on nuclear spin or hyperfine interaction)
using a beam of neutral atoms \cite{97:Hagley}.
The solid state proposal is to use 
low-mass charged particles (electrons) moving
in a solid to provide the coupling \cite{98:Kane}.

Elaboration on the above summary would enable us to see various strengths
and weaknesses of current
proposals. However, our purpose here is to seek
a method which appears to be in some sense natural, that is,
which makes use of physical principles which are naturally suited to the
task. We suggest that the natural, and possibly in the long term the best, choice
for the entity $\cal G$ which provides rapid controllable coupling, is light. 
Light is in any case the fundamental coupling between charges. It will
travel at the fastest possible speed between qubits, it 
can be made to appear and disappear at will and, perhaps most importantly,
it does not couple to extraneous electromagnetic fields in the computer,
which greatly reduces possible decoherence mechanisms. Furthermore,
photons provide a natural way to couple quantum information
out of the computer and down a quantum communication link.

To ensure there is no light when it is not wanted, we should use frequencies well
above the thermal spectrum at the temperature of the computer, but
otherwise the main consideration is ease and precision of manipulation
of the light (including the ability to select individual qubits).
This suggests near-infra-red or optical frequencies.
When we wish to couple resonantly to the hyperfine
splitting, which is in the microwave domain, we use Raman transitions.
Note that the typical frequency scale for hyperfine splitting, i.e. GHz,
is attractive because electronic
techology allows the most precise control in this frequency regime. This
is likely to remain true in the future, owing to basic properties of
electromagnetism and conduction in metals.

At this point in the argument,
we may envisage an ideal quantum computing system 
as based on an array of nuclear spins inside atoms, the spins
coupling to the electrons of their atoms, and the electrons coupling to
photons which ferry information around the computer,
appearing and disappearing at the behest of the controlling
machinery. The only remaining question is, how is the electron-photon coupling
to be both strong enough, and under sufficient control?

To achieve a strong enough coupling, the light must be confined in a small
volume, and to permit coherent coupling we
require a long photon storage time in the confining cavity, as well as
accurate positioning of the atoms.
These considerations will be addressed in section \ref{s_gate}.

It is possible to imagine that the atoms
might be held in place by any one of a number of methods, including attaching
them to long chain molecules or fabricating nanoscale structures to
hold them. However, the additional atoms and electrons which form the
body of any such structures will introduce new degrees of freedom which may
cause decoherence, or weaken the light-atom coupling. One possibility
is to situate the atoms on the surface, or perhaps under the surface,
of a highly transparent solid (see section \ref{s_design}).
In this paper we will concentrate on the case that the atoms
are held in an r.f. Paul trap (ion trap) or an optical dipole
force trap, and consider
the possibility of placing the atoms on a surface in the final section.

We note that whereas we have advocated using the nuclear
spin alone as quantum memory, current experiments designed to achieve quantum
information processing in ion traps are not operating in this regime.
In the work of Wineland {\em et al.}
\cite{95:Monroe,98:Turchette,00:Sackett}
with the beryllium ion the qubit involves both nuclear and electron spin, since
its energy level separation is a sum of hyperfine and first-order Zeeman
effects. A pair of electronic states (Coulomb interaction
with the nuclear charge) with first order Zeeman effect
is adopted in \cite{99:Roos,99:Nagerl,98:Hughes} and the electron spin alone
in \cite{98:Stevens}. We envisage
that all these experiments will contribute to the overall development
of the field, and it will be a relatively small step to adapt them
to hyperfine transitions with no electron spin component.

\section{Strengths of ion trap technology}  \label{s_iontrap}

Before discussing future prospects, we will highlight in this section
the strengths of current ion trap technology.
 
We consider the three requirements for quantum information processing,
which are quantum memory,
quantum logic gates, and measurement of quantum states. The primary
consideration for all of these is precision and reliability, simply
because we need the computer to work; we are willing to sacrifice 
both speed of operation, and ease of construction, if it makes the
difference between a computer which works and one which does
not\footnote{For a large computer this will, of course, only
be possible if the speed does not fall, nor the system size
increase, exponentially with the number of qubits.}.
 
Note that all the three requirements are equally significant. In particular,
measurement is at the heart of error
correction protocols \cite{98:Steane,99:SteaneB},
therefore it is a central consideration during the whole operation of
the computer, not merely at the final step where the computation
result is measured. For some purposes, it is sufficient that a
dissipation process can be applied to chosen quantum bits at chosen
times \cite{99:Aharonov}. The dissipation process forces the qubit to a known
final state, no matter what its initial state.
This can be easier to implement, so will be considered also. 

\subsection{Quantum memory and single-qubit gates}

We discussed the attractive features of nuclear spins for quantum computing
in section \ref{s_ideal}. Experiments with trapped
atoms and ions offer the most precise methods known for manipulation
of the nuclear spin, via the hyperfine interaction. Indeed, 
time and frequency standards throughout the world
are based on optical manipulation of
atoms trapped in high vacuum, and ion trap frequency standards now
rival those based on neutral atoms. This is the first advantage of
ion trap methods. From a practical point of view it means that the
quantum memory and single-qubit gates are, broadly
speaking, solved problems, in that we
can envisage trapped ions whose nuclear spin state is as accurately
preserved and manipulated as anything which current technology allows.

\subsection{Read-out}

The read-out is also, broadly speaking, a solved problem
for experiments with trapped atoms or ions. The measurement of
the hyperfine state can be carried out rapidly 
by the electron shelving (or `quantum jump')
method, which offers close to 100\% reliability
\cite{86:Nagourney,86:Sauter,86:Bergquist,98:Stevens,98:Turchette,99:Nagerl}.
The timescale for
such a measurement is set by the need to scatter a few
thousand photons on an allowed atomic transition, requiring
of order a few hundred $\mu$s.
For dissipation, we can use the method of optical pumping. Here,
only a few photons need to be scattered before we can be
confident the system has relaxed, so the time scale for controlled
dissipation is of order $0.1\;\mu$s.

A further point about measurement is significant:
as long as separate atoms or ions can be resolved by an
optical imaging system (implying a separation of at least
a few wavelengths) then they can be measured simultaneously.

The central problem for ion trap quantum computing is,
then, the question of implementing the 2- or more-bit logic gates.
This will be discussed in section \ref{s_gate}.

\subsection{Quantifying qubits, gates and entanglement}  

The standard way to quantify the complexity of
an algorithm on any computer, whether
quantum or classical, is to count the number of bits in the
memory and the number of 2-bit or 3-bit logic gates used. In the case
of quantum computing, it makes sense also to have a measure of 
the degree to which an algorithm involves highly non-classical effects.
A useful measure is to ask whether $n$-particle
entangled states, such as the ``Schr\"odinger cat'' state
$\ket{000 \cdots 0} + \ket{111 \cdots 1}$, can be produced.

So far no ion trap experiment has combined all the necessary features
to allow general processing on more than one ion. 
However, all the ingredients of general processing
have been demonstrated in separate
experiments\cite{95:Monroe,98:Turchette,99:Nagerl,99:Roos}, and
highly entangled states have been produced\cite{98:Turchette,00:Sackett}.

The definition of entanglement requires some comment. 
If two or more separate spin-half particles are in a
joint state $\ket{\phi}$,
then the existence of a non-zero overlap with an entangled state does not
necessarily
imply the presence of entanglement. For example, the overlap between the
4-qubit separable state $\ket{++++}$ (where
$\ket{+} = (\ket{0} + \ket{1})/\sqrt{2}$)
and the cat state $(\ket{0000} + \ket{1111})/\sqrt{2}$ is $1/8$.
Also, a superposition
$\ket{M=+3/2} + \ket{M=-3/2}$ of the two stretched states of a spin-$3/2$
particle is not entangled, even though it could be written
$\ket{0}\ket{0} + \ket{1}\ket{1}$ by a suitable choice of state labels.
The latter point is important: the mere fact that a state can be
written $\ket{0}\ket{0} + \ket{1}\ket{1}$ in some basis is {\em not}
enough to mean that it is entangled in any sense which is significant
to quantum information physics. 
A strict definition of the term ``entanglement'' would restrict its
use to refer only to degrees of freedom which could in principle be located
in separate spatial locations (so that entanglement-enhanced communication
could be realised), or which could be used to gain the
reduction in computational complexity offered by quantum computation
(compared to classical computation) for certain algorithms.
In that case the state of spin and
motion of an electron emerging from a Stern-Gerlach apparatus
would not be regarded as entangled.
However, it has become quite common
to broaden this strict definition slightly, so as to include the
case of ``entanglement'' between
the internal and motional degree of freedom of a single particle.

Such entanglement has been achieved between
the internal state of a single trapped
ion and its motional state, with a high degree of
precision and control, in at least two laboratories
\cite{96:Monroe,99:Roos}. A measure of the degree of entanglement is the size
of the Hilbert space in which coherent evolution is
demonstrated in the experiment. For example, the ``Schr\"odinger cat''
states realised in \cite{96:Monroe} involve a superposition of
coherent states. Each coherent state has a
Poissonian distribution over vibrational levels, characterised
by a parameter $\alpha$ which had the value $\alpha = 2.97 \pm 0.06$
in the experiments. The
mean vibrational quantum number $\langle n \rangle = | \alpha |^2
\simeq 9$, and standard deviation $\sigma_n \simeq |\alpha|
\simeq 3$.
The size of the motional Hilbert space in which coherent evolution must
take place in order to observe the interference is of order
$\log_2 ( \langle n \rangle + 1 + \sigma_n ) \simeq 3.7$ qubits.
Adding the internal degree of freedom, this is a 
``cat state'' of $4.7$ qubits.

A true multi-particle entanglement is very rare in physics,
and indeed for more than three particles it has only been achieved,
to our knowledge, in a single experiment. This is the
4-particle entanglement recently demonstrated in an ion trap experiment
by Sackett {\em et al.} \cite{00:Sackett}.

The strength of ion trap experiments which is underlined by this
achievement is that the generation of entanglement is under complete
experimental control: it is deterministic, rather than being the result
of a process which relies on an essentially random event (e.g. spontaneous
parametric down-conversion, or velocity selection of atoms from
a thermal source). This
is a significant distinction because the amount of
$n$-qubit entanglement produced by a random process
falls off exponentially with the number $n$ of qubits, and therefore
results in a system which cannot exhibit some of the essential
defining features of quantum computation, such as the breaking of
the classical hierarchy of complexity classes.
 
It is noteworthy that in all
experimental ``realisations'' of quantum algorithms so far
reported, the size of the apparatus, or the
duration of the experiment, has scaled with the number of qubits
required to define the problem
at the same rate or worse than a classical computer or
information channel would scale with the number of classical bits.

This does not mean that randomly
produced entanglement is uninteresting, since it can be used
to demonstrate some of the basic principles of quantum mechanics and
quantum information. However, one might draw an
analogy with the properties of light sources: a thermal
source, with a sufficiently narrow filter in front of it,
can produce radiation with just as narrow a bandwidth as
is available from a laser, but there remains a qualitative,
and practically significant, difference between thermal
radiation and laser radiation.
 
For light sources, a useful parameter which emphasizes that
bandwidth is not the only consideration is the number of
photons per mode. It would be useful to have a comparable
measure for entanglement, such as ``the number of singlets per
2-qubit Hilbert space'' (a singlet being the 2-qubit
entangled state $(\ket{01} - \ket{10})/\sqrt{2}$).
The difficulty in forming such a measure
is that the Hilbert space size, unlike the modes of a radiation
cavity, depends on which parts of the system we choose to focus
our attention on. For example, we may consider all the atoms
in a thermal beam, or just those selected by a velocity selector.
The least ambiguous measure is arguably that implicit in
\cite{98:Turchette}: we define the {\em entangling efficiency} 
$\epsilon$ to be
\begin{quote}
the probability that, starting from initial conditions of no entanglement, 
a singlet can be caused to be present in a predetermined Hilbert space
at a predetermined time.
\end{quote}
The predetermined Hilbert space means we indicate which systems
(eg atoms, spins, photons) will contain the singlet, without the
need to check by measuring them, and the predetermined time
means we decide beforehand at which moment we want the singlet,
without reference to the details of the experimental apparatus
(thus ruling out statements such as ``1 ms after detector D
clicks'', if we can't predict when detector D will click).
The purpose of the quantity defined is to enable us to assess rapidly
the slow-down to be expected when the same
apparatus is used to form 3-particle, 4-particle and
higher forms of entanglement. 

In parametric down-conversion experiments reported
to date \cite{99:Bouwmeester}, $\epsilon$
was of order  $10^{-4}$, and in cavity QED experiments
using a thermal atomic beam\cite{97:Hagley},
it was $\epsilon \simeq 3 \times 10^{-3}$.
For thermal ensembles such as those in current
liquid state NMR experiments it is
zero \cite{99:Braunstein}. A related quantity,
the entangling rate (number of successful singlet-generating 
runs per unit time) was approximately
8000 s$^{-1}$ and 2 s$^{-1}$ 
respectively for the down-conversion and CQED experiments. 

The first observation of an entangling efficiency of order 1
was in the experiment of Turchette {\em et. al.}
\cite{98:Turchette}. The internal state of two
trapped beryllium ions was driven to a singlet state with
reliability approximately 70\% (with entangling rate approximately
30000 s$^{-1}$). The recent report of
4-qubit entanglement arose from further 
work in the same laboratory \cite{00:Sackett}.
These remain the only demonstrations of a
high entangling efficiency in any area of physics.

\section{Experiments feasible in the short term}  \label{s_short}

We have noted that the main strengths of current ion trap experiments, compared
with other quantum information experiments, are that they allow rapid
and reliable measurement, and deterministic entanglement. The following
list concentrates on experiments which exploit these strengths, 
identifying goals which are either not realisable at all in other systems,
or for which the ion trap may be the system of choice.

A single trapped ion allows the experimental exploration of
two important avenues: the vibrational degrees of freedom,
and cavity QED \cite{95:Pellizzari,99:Ye,00:Pinkse}. The
interest of the vibrational degrees
of freedom is illustrated by the Schr\"odinger cat \cite{96:Monroe}
and environment engineering \cite{00:Myatt} experiments which further
our understanding and control of decoherence.

With 2 ions in the same trap, some standard quantum information ideas can
be demonstrated, such as the EPR experiment \cite{Bk:Peres},
``dense coding'' \cite{92:Bennett,96:Mattle}
and a simple ``algorithm'' such as Grover's search algorithm \cite{97:Grover}. Of
these, the EPR experiment is the most significant, since the detector
efficiency problem can be avoided \cite{Bk:Peres}; however the
close spacing of the ions makes impractical a test involving
space-like separated measurement processes. A demonstration of dense coding
would be the first time this idea had been implemented without
needing post-selection\cite{96:Mattle}, and hence allowing an unambiguous
increase in the capacity of the quantum channel to transmit
classical information. However since the `channel' involved only covers a
distance of some tens of microns in vaccum, it is of no practical use.

With only 2 qubits it is debatable whether the
most significant features of Grover's algorithm can be
demonstrated, but the algorithm would provide a useful way
of showing that quite general manipulations of a two-ion system
had been achieved.
 
With 3 ions three highly significant experiments could be done.
These are ``teleportation'' \cite{93:Bennett}, entanglement-enhanced
communication \cite{97:Cleve,00:SteaneA}, and quantum error
correction \cite{96:SteaneB}.
In addition, a thorough (though still very simple) demonstration 
of Grover's algorithm would be possible.

Quantum teleportation is significiant not only in the context
of quantum communication, but also as an essential ingredient
of fault-tolerant quantum processing\cite{99:SteaneB,99:Gottesman}.
A reliable teleportation experiment within a small quantum processor
would therefore be a significant development.

The most accessible example of
entanglement-enhanced communication is the ``Guess my number'' 
protocol \cite{00:SteaneA}, in which three parties use shared
entanglement and classical communication to learn the answer
to a simple mathematical problem.
In order to obtain a result which breaks
the classical limits on communication, an experiment of
overall reliability above 50\% is needed. 

The simplest example of quantum error correction requires
3 qubits, which are used to protect a single logical qubit
against a restricted class of errors \cite{96:SteaneB}. The set of correctable
errors could be, for example, phase errors on single bits.
The most striking result is obtained, however, if the
errors are not merely unitary precession of the qubits
themselves, but non-unitary relaxation processes where information leaks
away into the environment. For example, optical
pumping could be used to cause a relaxation of one qubit,
where, after tracing over the environmental degrees of freedom,
the qubit has `collapsed' into a mixed state, with density matrix
$P_0 \ket{0}\bra{0} + P_1 \ket{1}\bra{1}$. After this, 
the correction network is applied, and it 
would still recover the exact encoded state in the three qubits.
Furthermore, the process of random error followed by correction, could be
repeated many times on the same encoded state.

This process is remarkable from several points of view. After it is
repeated a few times, the environment would have had a chance to
``measure'' {\em all} the qubits in the processor, thus causing, one
would think, a large perturbation to the state, and yet the qubit
of information is perfectly preserved. Alternatively, if we
drive optical pumping continuously but weakly on all the qubits, then
the loss of fidelity of the qubits is linear with time, for small times,
while after correction it becomes quadratic with time, therefore
allowing the Zeno effect to be implemented in the case of a relaxation
process \cite{57:Khalfin,61:Winter,77:Misra,97:Wilkinson}.

The 3-qubit quantum error correction has been investigated in
an NMR experiment \cite{98:Laflamme}. Some aspects of the expected
behaviour were demonstrated, but owing to the limitations
of the pseudo-pure state method, a genuine error correction
was not available, since the entropy could not be extracted
from the system. This is seen most clearly in two aspects of
the experiment. First, in encoding from one qubit into three,
the signal size fell by a factor 8, and no subsequent error
correction can make up for the increased sensitivity to errors in this
situation. Secondly, it was not possible to apply correction
repetitively. 
 
Note that all the experiments we have listed for three ions rely
on the ability to perform not just unitary processing
operations, but also strong measurements of one or more
chosen qubits.
Successful realisation of the ``Guess my number''
or the repeatable quantum error correction protocols
would be landmarks in quantum information science.

Obviously, there are more and more experiments which are
possible as the number of qubits increases, even 
before useful quantum computation becomes possible.

\section{Logic gate methods in ion traps}  \label{s_gate}

The first general method proposed to implement
quantum logic gates between trapped ions was that discovered
by Cirac and Zoller \cite{95:Cirac}. This is based on using the motional
degree of freedom to ferry quantum information from one ion
to another. All ion trap quantum processing
experiments so far have been based on this idea.
A significant further insight was provided by 
M{\o}lmer and S{\o}rensen \cite{99:Molmer,99:Sorensen}
who showed how to make better use of the motional degree of freedom,
and the recent experiments of \cite{00:Sackett} are based on these further
insights.

Another method to couple separate atoms or ions coherently is
to use light to ferry the information around, using the proposal
of \cite{95:Pellizzari} based on concepts in
cavity quantum electrodynamics (CQED).

In this section we will compare the two methods---motional
and photon-based gates. We assume some familiarity with both
methods on the part of the reader.
At present the motional methods are
easier to achieve experimentally, but the CQED methods
allow, in principle, higher gate rates, and also 
quantum communication
between separate ion- or atom-traps. 

There is a subtlety regarding the hyperfine interaction and the
optical transitions involved in these methods. The electric dipole
optical transitions which we will use do not couple directly to
the nuclear spin. In the motional coupling, this implies that the
change of
internal state of the ion must involve the electronic wavefuntion,
so it is not purely a nuclear spin rotation. As a result,
the relevant hyperfine levels will typically have a
first-order Zeeman effect. In the CQED method,
4 states in the ground hyperfine manifold of a single ion are used.
In either case, during the action of the gate, the quantum information
is stored in electronic not nuclear degrees of freedom. However,
the quantum memory can remain a wholly nuclear spin system: 
we swap quantum information between Zeeman levels $\ket{M_F}$
just before and just after each gate, by driving a
Raman transition in the internal state of the ions involved in the gate.
We assume this transition can
be fast compared to the gate operations to be discussed.

\subsection{Motional coupling}  \label{s_motional}

The Cirac-Zoller method to implement 2-qubit gates such as
`controlled not' between separate ions is to couple the internal
state of chosen ions to the vibrational degree of freedom,
by driving Rabi flopping on a vibrational sideband of the
atomic transition. The phenomena which cause the main limitations
of this method are relaxation and/or heating, and off-resonant driving of
unwanted transitions.

\subsubsection{Relaxation}

There are two main sources of relaxation in the ion trap. These
are the spontaneous decay of excited states of the ion, and the heating
or relaxation of the vibrational degree of freedom. To minimise
the effects of these, a compromise between fast and slow operation
of the processor is needed.

\begin{figure}[h]
\centerline{\resizebox{!}{8 cm}{\includegraphics{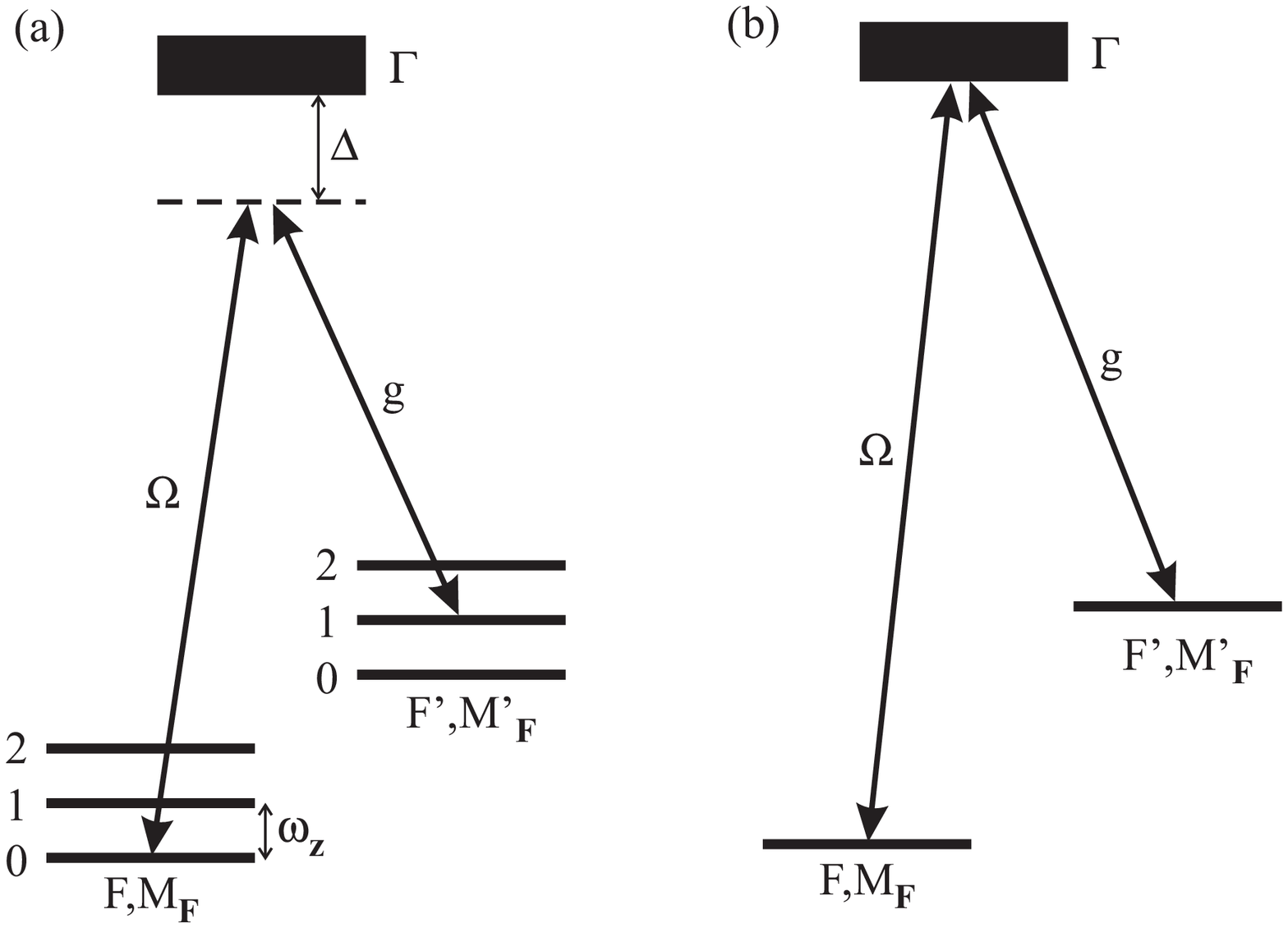}}}
\rule{0pt}{24 pt} 
\caption{Energy levels and notation for (a) motional gates, (b) CQED
optical gates.}
\label{f_levels}
\end{figure}

The quantum gates between ions involve the excitation of the
motional degree of freedom, so we 
consider driving the first red vibrational sideband of a
resonant Raman transition between hyperfine levels in a trapped
ion \cite{97:SteaneB}, see figure \ref{f_levels}.
The Raman transition is driven by a pair of lasers detuned
by $\Delta \gg \Gamma$ from an allowed single-photon transition
whose natural width is $\Gamma$ (full width half
maximum in angular frequency units).
The Rabi frequencies of the relevant single-photon transitions
are $\Omega$ and $g = \eta \Omega$, where $\eta$ is the
Lamb-Dicke parameter. In this situation the pair of hyperfine levels
connected by the Raman transition form an effective two-level
system; the two-level transition has
effective Rabi frequency $\Omega_{\rm eff} = \Omega g / 2 \Delta$.
A two-ion gate such as a state-swapping operation requires
two $\pi$ pulses on a vibrational sideband, so the total time
is $T=2 \pi / \Omega_{\rm eff}$. During
each pulse the mean population of the unstable excited state of
the ion is $\Omega^2 / 4 \Delta^2$ (assuming $g \ll \Omega$), therefore
the mean number of photons scattered is
\beq
p_1 = \frac{\Omega^2}{4 \Delta^2} \Gamma T
= \frac{\pi \Gamma \Omega}{\Delta g}.        \label{p1}
\eeq
We assume that at all times the vibrational state of the ion suffers
a non-unitary heating process, characterised by a rate $\kappa$
which is the rate of heating (or relaxation)
from one vibrational state to an orthogonal one. Therefore the
probability of relaxation by this process, during the two pulses,
is
\beq
p_2 = \kappa T = \frac{4 \pi \kappa \Delta}{\Omega g}. \label{p2}
\eeq
The total probability of failure is 
\beq
p = p_1 + p_2 = \frac{\pi \Gamma}{g} \frac{\Omega}{\Delta}
+ \frac{4 \pi \kappa}{g} \frac{\Delta}{\Omega}.
\eeq
In this equation $\Gamma$ is constant for a given atom,
and $\kappa$ is characteristic of a given experimental apparatus,
while $\Omega/\Delta$ can be adjusted to minimise $p$. This minimisation
gives $\Omega/\Delta = 2 ({\kappa}/{\Gamma})^{1/2}$, and
\beq
p_{\rm min} = 4 \pi \sqrt{ \frac{\kappa \Gamma}{g^2}\,}, \;\;\;\;\;\;
\frac{1}{T} = \frac{g^2}{\Gamma} \frac{p_{\rm min}}{8 \pi^2}.      \label{Tr} 
\eeq

\subsubsection{Off-resonant coupling}

The vibrational levels in an ion trap are typically closely spaced
compared to all other energy level separations in the system,
so the transitions driven off-resonantly are primarily the
carrier transitions (those which don't change the vibrational state),
which are off-resonant by the vibrational
frequency $\omega_z$. This problem is studied in detail in
\cite{00:SteaneB}. The conclusion is that after two $\pi$ pulses,
the amount of population which
leaks into unwanted states due to off-resonant coupling is
\beq
p_3 = \frac{\Omega^2_{\rm eff,0}}{\omega_z^2}
= \left( \frac{\Omega^2}{2 \Delta} \right)^2 \frac{1}{\omega_z^2}.
\label{p3}
\eeq
where $\Omega_{\rm eff,0}$ is the Rabi frequency for carrier transitions. 

\subsubsection{Discussion}

Of the processes we have considered which limit the motional
coupling, the atomic relaxation $\Gamma$ and the off-resonant
excitation are intrinsic to the physics of the system, while the
motional heating $\kappa$ could in principle be made arbitrarily
small (in recent experiments values of $\kappa / \omega_z$
as low as $2 \times 10^{-7}$ for a single ion \cite{99:Roos}
and $2 \times 10^{-6}$ for the stretch mode of two ions \cite{98:King}
were reported.)
Therefore equations (\ref{p1}) and (\ref{p3}) give
the main limitations. The gate rate is limited by (\ref{p3})
since for a given value of $\Omega_{\rm eff,0}/\omega_z$ 
we can make $p_1 \ll p_3$ by increasing the laser intensity
and its detuning $\Delta$ (until the detuning becomes comparable
to further energy-level separations in the ion, such as the
fine structure).

In the case that motional heating is significant, the
M{\o}lmer-S{\o}rensen approach may be advantageous, since it permits
gates of high fidelity in the presence of motional heating
(at the expense of reduced gate rate). However, in the limit
of small $\kappa$, the gate rate produced by this method in its
standard form is limited by off-resonant excitation and is the same
as that given in equation (\ref{p3}).
 
In conclusion, the gate rate at given failure probability $p$
for a 2-qubit swap gate via the motional state is (from (\ref{p3}))
\beq
\frac{1}{T} = p^{1/2} \eta \frac{\omega_z}{2 \pi},
\label{Toffr}
\eeq
assuming $\kappa \ll p / T$ and $\Delta \gg \pi \Gamma/\eta p$, and
using either the Cirac-Zoller or the M{\o}lmer-S{\o}rensen methods.

It is notable that this limit, imposed by off-resonant carrier transtions,
could be exceeded by exciting the ion in the node of a laser
standing wave \cite{95:Cirac,98:James}. That is technically very difficult, but it
shows that the physics of the system can allow a faster switching rate.
Recently, less demanding methods to gain such a faster rate have been proposed.
The M{\o}lmer-S{\o}rensen approach can probably be made to produce
faster gates than (\ref{Toffr}) by a careful choice of
parameters \cite{00:Sorensen}, and
recently a new approach has been put forward in which such a speed up
is thoroughly analysed \cite{00:Jonathan}. The latter uses a
light-shift-induced resonance, yielding a gate rate $\eta \omega_z/2\pi$
and gates of fidelity approximately $1-\eta^2/2$. Therefore
the speed increase compared to eq. (\ref{Toffr}) is significant, of
order $1/\eta$.

\subsection{Photon-based coupling} \label{s_photon}

We will now consider coupling qubits via the excitation of a mode
of a high-finesse optical cavity.

We will assume an allowed electric dipole transition,
so the Rabi frequency describing the coupling is $g = E d / \hbar$ where $E$
is the electric field of the light, and $d$ is the electric dipole matrix
element.
We will use this coupling to exchange quantum information between an
atom and a light field by absorption or emission of single photons, therefore
we are interested in the value of $g$ when the electric field is that of a
single photon. If the photon has angular frequency $\omega$ and
occupies a mode of volume $V$, then its
energy is $ \hbar \omega=\epsilon_0 E^2 V /2$, hence
\beq
g = d \sqrt{\frac{2 \omega}{\epsilon_0 \hbar V}}      \label{g}
\eeq
The strongest electric dipole matrix elements in atoms are all of order
$e a_0$\
where $e$ is the charge on the electron and $a_0$ is the Bohr radius.
The spontaneous
decay rate $\Gamma$ of an atom on a strong transition varies as $1/\lambda^3$:
\beq
\Gamma = \frac{\omega^3 d^2}{3 \pi \epsilon_0 \hbar c^3}     \label{Gamma}
\eeq
so for
$g \gg \Gamma$ the long wavelength region is best. However, we will need the
logic gates to be fast, setting a premium on large $g$, hence small wavelengths.
The other major source of decoherence is decay of the photon mode owing to the
finite finesse of the cavity which contains it. This decay rate is
\beq
\kappa = \frac{c \pi}{ {\cal F} L }    \label{kappa}
\eeq
for a cavity of length $L$ and finesse $\cal F$.

\subsubsection{Dark state and adiabatic passage}

Since we need very precise gates, there is interest in any
method to implement them which
has reduced sensitivity to the relaxation of the atom and the cavity photon.
Such a method is the adiabatic passage, as described in \cite{95:Pellizzari}. Briefly,
a state-swapping operation is carried out between any two atoms in the cavity
by shining laser pulses on the two atoms. The pulses are not exactly
simultaneous, but overlap in time, coupling ground
states $\ket{a}_i$ to each atom's
excited state with Rabi frequencies $\Omega_i$, for atoms $i=1,2$ respectively.
The cavity mode produces strong coupling between the excited state and
a metastable level $\ket{b}_i$ simultaneously
for both atoms. In our case,
$\ket{b}_i = \ket{F', M'_F}_i$ is in the ground state manifold, separated
from $\ket{a}_i = \ket{F,M_F}_i$
by the hyperfine interaction.
The system of two atoms plus cavity photon exhibits the phenomenon of dark
states, i.e. superpositions of states which by quantum interference are
decoupled from the excited states. For up to one photon in the cavity, there
are two dark states \cite{95:Pellizzari},
\begin{eqnarray}
\ket{D_0} &=& \ket{b,b,0} \equiv \ket{b}_1 \ket{b}_2 \ket{0}_c, \label{dark} \\
\ket{D_1} &\propto& \Omega_1 g \ket{b,a,0} + \Omega_2 g \ket{a,b,0}
- \Omega_1 \Omega_2 \ket{b,b,1}.    \nonumber
\end{eqnarray}
The swap operation carries one qubit of information from one atom 1 into atom
2. A general operation such as controlled-not is then carried out within
atom 2 using four of its states (2 Zeeman components of each of 2 hyperfine
levels), then the information is swapped back. Any other atoms in the cavity
do not participate because they are not illuminated by the laser pulses,
and they are in the states $\ket{a}$ which are not coupled to the cavity
photon. To ensure the off-resonant coupling is sufficiently small, we will
require the hyperfine splitting to be much larger
than $\Omega,\;g$.

The method of adiabatic passage is limited by two considerations.
First we need to preserve adiabaticity, and second we need to avoid
populating states which suffer non-unitary relaxation. Numerical
solution of the master equation for the complete system is discussed
in \cite{95:Pellizzari}. Here we will make rough estimates in order to identify
the best operating regime, for given
parameters $g$, $\Gamma$, $\kappa$.

To preserve adiabaticity the rate of change of the conditions must
be slow compared to the frequency separation between the state 
we wish the system to remain in (here, the dark state) and any
other state (here, the nearest bright state). For example,
if the frequency separation $\omega_{ij}$ and the rate of change
of the state are constant in time, then the probability to make an
(unwanted) transition from the desired state $\ket{i}$ to some
other state $\ket{j}$ after time $t$ is \cite{Bk:Messiah}
\beq
P_{i\rightarrow j} \simeq \left| \frac{dH_{ji}}{dt}
\frac{1}{\hbar \omega_{ji}^2}
\right|^2 2 \left( 1 - \cos \omega_{ji} t \right).  \label{Pij}
\eeq
In our case, we will assume the laser pulses have the form
$\Omega_2(t) = \{0, \Omega t/T, \Omega\}$ for
$\{ t\le0, 0<t<T, t \ge T\}$ respectively, and
$\Omega_1 = \Omega - \Omega_2$, therefore $dH_{ji}/dt = \hbar \Omega/T$.
The frequency separation between
the dark state and the nearest bright state is a complicated function
of the Rabi frequencies, which we simplify
to $\omega_{ji}(t) \simeq (\Omega_1^2 + \Omega_2^2)^{1/2}$
(it  will emerge that for the cavities we will consider, we will
need $\Omega \ll g$ in order to minimise the effects of relaxation of the
cavity mode). For $P_{i\rightarrow j} \ll 1$
the oscillating term in (\ref{Pij}) averages to zero during any
interval of time small compared to $T$. We
find the probability to make a transition
out of the dark state by using the derivative of (\ref{Pij})
with respect to time, and then integrating from $t=0$ to $T$,
to obtain
\beq
p_1 \simeq \frac{4}{T^2 \Omega^2}.
\eeq

During the adiabatic passage, there is population in the
state $\ket{b}_1\ket{b}_2\ket{1}$ which decays at a
rate $\kappa$. We model this by assuming that at any time $t$
the system is in a mixture containing 
dark state population $1- \int P_{bb1} \kappa dt$,
and non-dark state population $\int P_{bb1} \kappa dt$, where
$P_{bb1}$ is the population of the 
$\ket{b}_1\ket{b}_2 \ket{1}$ component in $\ket{D_1}$, which,
from (\ref{dark}), is approximately
$\Omega_1^2 \Omega_2^2/g^2(\Omega_1^2 + \Omega_2^2)$,
for $\Omega \ll g$. The non-dark part of this mixture will
be strongly coupled to the environment by photon scattering, but
the dark part is unaffected by any process involving $\Gamma$.
Therefore the net loss of fidelity is given by the integral
of $P_{bb1}\kappa$ over the switching time, which is
\beq
p_2 \simeq \frac{\Omega^2}{2 g^2} \kappa T.
\eeq
The net failure probability of the gate is $p = p_1 + p_2$. We choose
$T$ to minimise $p$, and then express the gate rate in terms of
$p$ and the coupling parameters, obtaining:
\beq
\frac{1}{T} \simeq \frac{1}{9} \, p^{2} 
\frac{g^2}{\kappa}.   \label{rateAP}
\eeq
Note that we can obtain a gate of as high fidelity as we wish (against
decoherence by the mechanisms we have considered) by slowing down
the processor. The laser intensity must be chosen to give
$\Omega = g^2 (p/3)^{3/2}/ \kappa$ to meet the conditions
of maximum fidelity. We used the approximation $\Omega \ll g$, which
will be valid for this choice of $\Omega$ when $g/\kappa \ll (3/p)^{3/2}$.

\subsubsection{Rabi flopping}  \label{s_rabi}

The method of adiabatic passage suffers from the problem that
the gate rate given in equation (\ref{rateAP}) scales as the
square of the failure probability, in contrast to the better
scaling properties of equations (\ref{Tr}) and (\ref{Toffr}).
To avoid this poor scaling,
the single-photon mode can be used in a way
analogous to that adopted for the vibrational mode
in section \ref{s_motional}:
instead of adiabatic passage in a dark state, we drive Rabi
flopping on a chosen atom $A$, using a $\pi$ pulse to place a
single photon in the cavity mode, and then a similar pulse on
another atom $B$ swaps the information from the cavity mode
into atom $B$.

The analysis of this process is exactly as in equations
(\ref{p1})-(\ref{Tr}) except that now $g$ is the coupling between
atom and cavity mode, $\kappa$ is the relaxation rate of the
cavity mode, and since the cavity mode only decays
(it does not heat), the relaxation probability given in equation
(\ref{p2}) is halved, so we replace $\kappa$ by $\kappa/2$ in
equations (\ref{p2})-(\ref{Tr}), obtaining
\beq
p = 2 \pi \sqrt{ \frac{2 \kappa \Gamma}{g^2}\,}, \;\;\;\;
\frac{1}{T} = \frac{g^2}{\Gamma} \frac{p}{8 \pi^2}.      \label{TPr} 
\eeq
The gate rate is now limited by 
\beq
\frac{g^2}{\Gamma} = \frac{3 c \lambda^2}{2 \pi V},   \label{g2G}
\eeq
where we have used (\ref{g}) and (\ref{Gamma}). From general physical
principles, the minimum cavity mode volume $V$ might be expected to
scale with wavelength as $\lambda^3$, which would imply the processor
runs faster at shorter wavelengths. In practice the cavity
dimensions are also limited by technological considerations.

The state of the art for high finesse Fabry Perot
cavities in the optical domain
is indicated by \cite{99:Ye} (see also \cite{00:Pinkse}).
A pair of mirrors of radius of curvature
$10$ cm gave a finesse ${\cal F} = 4.2 \times 10^5$
at $\lambda = 852$ nm, and was used to form a cavity of
length $L = 44.6\micron$, Gaussian mode
waist $w_0 = 20\micron$, hence
cavity field decay rate $\kappa/(2 \pi) = 8$ MHz. The mode
volume $V = L w_0^2$ yields $g/(2 \pi) = 70$ MHz for
coupling to the D$_2$ line of atomic caesium, linewidth
$\Gamma/(2 \pi) = 5.3$ MHz. Equation (\ref{TPr}) then gives
$p \simeq 0.8$. 

Another important type of optical cavity is provided by the whispering
gallery modes of silica microspheres \cite{89:Braginsky,98:Ilchenko}.
Mabuchi and Kimble \cite{94:Mabuchi}
give the theory of coupling between the whispering gallery mode and an
atom positioned at or near the surface of the sphere. On the surface
of a $50\micron$ radius sphere, for example, the coupling to the D$_2$
line of neutral caesium atoms ($\lambda = 852$ nm)
is $g/\Gamma \simeq 6$ and for quality
factor $Q \simeq 2 \times 10^9$ which has been reported
\cite{93:Collot},
$g/\kappa \simeq 174$, giving $p \simeq 0.3$.

Combining (\ref{kappa}), (\ref{g2G}) and (\ref{TPr}) we obtain
$p = (4 \pi^2 w / \lambda) (3 {\cal F})^{-1/2}$, where $w$
is the average diameter of the mode, so that $V = L w^2$. This
implies that it will remain very difficult to obtain $p \ll 1$
for a considerable time,
since great improvements in finesse will be needed, as well
as a reduction in the mirror or sphere radius of curvature (to reduce
the mode volume). 

\subsubsection{Discussion}

The advantage of the adiabatic passage method is that it allows
precise gates even in the presence of relaxation of both the atomic
excited state and the cavity mode. However, it is generally
true that adiabatic methods achieve their greater degree
of noise tolerance at the expense of processor speed.
In the present case, if a cavity of
sufficiently low decay rate can be built, then the Rabi flopping
method may be preferable in that it will be faster 
in the limit of small $p$, and may perform better against
further considerations,
such as driving of off-resonant transitions. 

Since the physics of the vibration of an ion string is
similar to that of the excitation of a cavity mode, it
ought to be possible to apply a method analogous to the
M{\o}lmer-S{\o}rensen one \cite{99:Molmer,00:Sorensen}
to the case of photon coupling. The essential
result of the M{\o}lmer S{\o}rensen method is that the sensitivity
to relaxation of the degree of freedom providing coupling
($\kappa$) is reduced by a factor $M$, while the gate time gets
longer by the factor $M$. Examining (\ref{TPr}) we see that in
order to halve $p$, we would need to multiply the factor
$M$ by four, increasing the gate time by a factor 4. The
gate rate thus scales again as $p^2$,
as it does in the adiabatic passage method.

\subsection{Gate time per ion}  \label{s_gatetime}

In the cases both of motional coupling and of photon coupling,
the gate rate is reduced when the number of ions $N$ in
the trap increases. For the motional coupling, this is
partly because the ion string gets heavier and so has a reduced
recoil frequency, and partly because if we wish to allow
individual ions to be resolved (for single-ion addressing)
the trap confinement must be reduced as more ions are added.
For the photon coupling,
the slow-down arises because the mode volume must be large
enough to enclose all the ions, therefore reducing $g$.

We will assume that the ions are in a linear string, or else
a rectangular array, each separated from its neighbours
by $s = 5 \lambda$. This means that if individual addressing
is achieved by directing separate Gaussian laser beams on
each ion, then the cross-talk between ions would be at the
level $10^{-4}$ if each beam had a waist $w = s / (\log 100)^{1/2}
\simeq 2.3\, \lambda$. Such a beam can be produced by optics
of modest numerical aperture. An alternative way to achieve
the individual addressing is discussed by Leibfried \cite{99:Leibfried}.

The scaling with $N$ of the motional coupling using the Cirac-Zoller
method is discussed in
\cite{97:SteaneB,98:James,00:SteaneB}. If we require the closest
ions in a string to be
separated by at least $s = s(\lambda)$, then the gate
time increases approximately as $N^{0.93}$. Approximating
this as proportional to $N$, and adopting the
breathing mode (vibrational frequency $\sqrt{3}$ times that
of the centre of mass mode) for the gates, we obtain 
from equation (\ref{Toffr}) a gate
time per ion which depends only on the failure probability $p$
and properties of the ion such as its mass and recoil frequency.
For candidate ions such as beryllium and calcium, this time
is of order $2$ and $10\;\mu$s respectively for $p=0.01$, when
$s = 5 \lambda$.

The faster method of \cite{00:Jonathan} produces a gate rate 
$\eta \omega_z/2\pi$ limited
through the limit on $\omega_z$ imposed by the need to keep ions
separated by $s$. We take $N=140$ as an example, which will be useful
in section \ref{s_design}. For the 313 nm transition in the
beryllium ion the requirement $s=5 \lambda$ 
leads to $\omega_z/(2 \pi) = 418$ kHz, hence
$\eta = 0.088$ and gate rate 37 kHz. The 397 nm transition in the
calcium ion gives $\omega_z/(2 \pi) = 141$ kHz, $\eta = 0.056$
and gate rate 8 kHz. Expressed as a gate time per ion, these examples
are $0.2$ and $0.9\;\mu$s respectively, with gate failure probabilities
of order $p \simeq \eta^2/2 \simeq 0.004$ and $0.0016$ respectively.

\begin{figure}[h]
\centerline{\resizebox{!}{8 cm}{\includegraphics{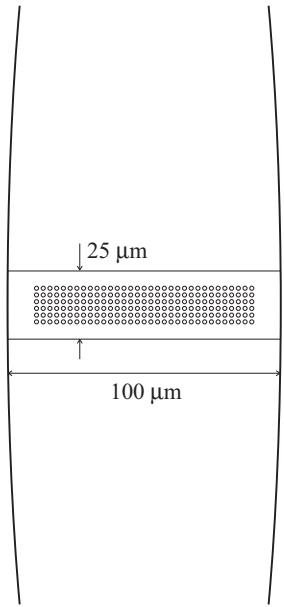}}}
\rule{0pt}{24 pt} 
\caption{Optical cavity for light-based coupling of 200 qubits, with parameters
appropriate to the barium ion. The qubits are shown in a rectangular array
spaced by $5\lambda$, which is possible if they are in fact neutral atoms in
an optical lattice. In an ion trap they would arrange themselves in another
pattern, but the essential features are unchanged (the ions can be made to
lie in a plane for a sufficiently elliptical trapping potential).}
\label{f_cav}
\end{figure}

For photon coupling we take as an example the cavity used
in \cite{99:Ye} whose properties are summarised in section \ref{s_rabi},
with two changes: we reduce the mirror radius of curvature
by a fifth (to $2$ cm), and we assume mirrors could be polished 
to provide the same finesse at half the wavelength.
To be precise, we choose $\lambda = 493$ nm which is appropriate
for the barium ion. This ion can be readily laser cooled, and
has the right kind of hyperfine structure for the adiabatic
passage approach. The cavity
mirrors are placed $L = 100 \micron$ apart, yielding a
mode waist $w_0 = 12.5 \micron$ and $\kappa = (2 \pi) 3.6$ MHz.
The mode can therefore accommodate about 200 ions and
equation (\ref{rateAP}) gives a gate time per ion of
$66$ ns for $p=0.01$ (using $\Gamma/(2 \pi) = 11$ MHz
for the $D$ lines in the barium ion,
we calculate $g/(2 \pi) = 62$ MHz). The cavity is illustrated in
figure \ref{f_cav}.

The conclusion is that for optical cavities which are
currently accessible or which may be expected in the near future,
the CQED and motional methods have similar speeds (which one
is faster will depend on $p$). It would be a lot harder to build
the combined optical/atomic system compared to an ion trap alone.
However, in principle the optical method could be
much faster if suitable cavities could be made, and one possibility
for this is a silica microsphere cavity. 

\section{Design for a quantum computer}  \label{s_design}

In view of the large number of technical problems still to be
investigated in the laboratory, it is premature to try to design
or build a large quantum computer. However, by sketching the
main features of a possible design, we can learn about the issues,
and identify avenues for further investigation. 

We aim to sketch a design for a quantum computer which could perform algorithms
requiring $10^6$ Toffoli (controlled-controlled-not)
gates on $100$ logical qubits. These numbers
are chosen on the basis that about 100 qubits are likely to be needed
to allow computations which could not be done on a classical computer. For
example, the input to the computation might require 20 qubits, and the
further 80 are needed as workspace. A problem needing less than 20 qubits
input can probably be solved more easily on a large classical computer. 
It is important to count Toffoli (or equivalent) gates, not 2-qubit gates
such as controlled-not,
because Toffoli gates are a major component in any quantum algorithm which
cannot be efficiently simulated classically, and, in contrast to
controlled-not, it is non-trivial
to implement them in a fault-tolerant
manner \cite{98:Preskill,99:Gottesman}. We take $10^6$ gates since
Shor's algorithm requires $O(k^3)$ gates for a $k$-qubit problem,
and a similar scaling is likely to be involved for any useful algorithm.

The computer will rely on fault-tolerant
methods and quantum error correction. To be specific, we will adopt the methods
described in \cite{99:SteaneB}: the quantum computer consists of blocks of 127 physical
qubits, where each block encodes 29 logical qubits in a 7-error-correcting
BCH code, and each data block is accompanied by several ancilliary blocks.
We place each block in a separate processor, and link processors together
by CQED and optical fibre
methods \cite{97:Cirac,97:Pellizzari,97:vanEnkA,98:Sorensen,98:Cirac}. In order to allow low-level
error correction methods in addition to those acting at the level of the
encoded blocks, each processor will contain 13 extra physical qubits, making
140 in all. These also serve to implement communication protocols between
processors, and for other tasks such as probing the local magnetic field.

\begin{figure}[h]
\centerline{\resizebox{!}{8 cm}{\includegraphics{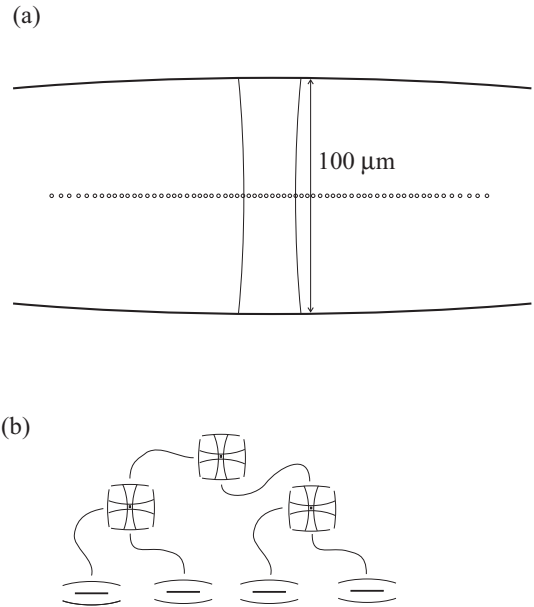}}}
\rule{0pt}{24 pt} 
\caption{Ion trap and cavity QED-based quantum computer. (a) A single processor
consists of 140 calcium ions in a linear trap, with an optical cavity around it. 
The cavity mirrors are separated by $100\micron$ and have radius of
curvature 2 cm; the mode waist is $w_0 = 11\micron$. Processing within the
trap is by motional methods; the optical cavity is used to implement gates
between ions in different traps.
(b) Processors are connected by optical fibres, and route switching is via
small traps with two cavity modes each, coupled into different
optical fibres.}
\label{f_trap}
\end{figure}

We will calculate the processor speed for two designs. The first is
a linear ion trap containing calcium ions as described in section
\ref{s_gatetime}. The gates between ions in a given trap
use the vibrational motion, and we 
adopt the fast gates offered by the lightshift-based
concept of Jonathan {\em et al.} \cite{00:Jonathan} or
by other methods which can be tailored in a similar fashion.
To network between processors each ion trap has around it a
Fabry-Perot optical cavity, with a mode shape overlapping several ions
at the centre of the trap (see figure \ref{f_trap}). Note that this
cavity should have parameters optimised for quantum communication between
traps \cite{97:Cirac,97:Pellizzari,97:vanEnkA,98:Sorensen},
not for the optical gates discussed in section \ref{s_photon}.
The relevant transitions in calcium have wavelengths around 400 nm.
We will also consider a more speculative
idea: an all-optical method involving no ion traps. Instead,
each processor is a single
silica microsphere, with 140 caesium atoms positioned around the
circumference of the sphere, either on the silica surface or trapped
near it by a dipole force optical trap \cite{94:Mabuchi}. The coupling between
spheres is by further optical cavities whose design is left unspecified
(it is not easy to see how to design them). The characteristics of the
sphere are as given in section \ref{s_rabi}, except that to accommodate
140 atoms spaced by $5 \lambda$ we require a sphere of radius $63\micron$
(for this calculation we use the wavelength in silica, since the atoms
can be addressed by directing laser beams through the sphere). This
reduces $g$ and $\kappa$ by a factor $50/63 \simeq 0.8$ from the values
quoted in section \ref{s_rabi}.

In \cite{99:SteaneB} it was assumed that multiple
controlled-not operations, in which there is one control qubit and many
target qubits, could be performed in a single time-step. This will not be
assumed possible here (but note the comments in \cite{00:Sorensen}), so
we must modify the results of \cite{99:SteaneB} accordingly.
The preparation of ancillas is slowed down,
which will increase the effect of memory noise. In order to reduce this
problem, we provide the computer with more ancilla blocks. It can
then prepare them in parallel, but staggered in time, so that enough
ancillas are always available when they are needed. Providing 40 ancillas
per data block (instead of 4 as in \cite{99:SteaneB}) reduces the memory noise
requirements by an order of magnitude. The whole computer then
needs 138 ion traps and associated optical cavities for the data
and ancilla blocks. Further optical cavities, with a few ions in each, may be
useful for switching information paths, so that each block
can communicate with most other blocks. This brings the total
number of ion traps or microspheres and associated cavities to around 200.
Using an analysis along the lines of that in \cite{99:SteaneB,97:SteaneC}, we
find that the quantum algorithm can be stabilised as long as the failure
rate per gate is $\gamma \simeq 10^{-4}$, and the
memory noise $\sim 10^{-6}$ per bit per timestep.

To obtain the gate failure probability $10^{-4}$ it is unlikely
that the best policy is to use the methods of section \ref{s_gate}
alone. Instead, we will assume we can implement a
low-level error correction tailored to the physical error process,
as for example in \cite{96:Cirac,97:vanEnkA,97:vanEnkB}. We assume the gates of
precision $10^{-4}$ can be built from gates of
precision $2 \times 10^{-3}$ at a cost of a factor 10 in
speed\footnote{These ratios of speeds and failure probabilities
are typical for a first-order (``single error'')
correction process which can use reliable measurements.}.
For the ion trap, the gate time is therefore approximately 
$10 / 8000 = 1250\;\mu$s, using the figures given in
section \ref{s_gatetime}.
The trap would use an axial vibrational frequency of 418 kHz
for the centre of mass mode, and frequencies above 20 MHz
for radial confinement.
Vibrational heating would need to be at the level
$\kappa / \omega_z < 10^{-6}$, which is within current
achievements. For the microsphere the gate time is
$10 \times 35 = 350\;\mu$s (from (\ref{rateAP}) with
$g = 4.8 \Gamma = 160$ MHz, $\kappa \simeq (g/170) \simeq 0.9$ MHz). 

To prepare and verify an ancilla takes $\sim 5000$ time steps
\cite{99:SteaneB,97:SteaneC}, but
since we prepare ten in parallel for each one we need, the
delay before the next one is available is approximately 500 time steps.
We envisage that this time is also sufficient to carry out one 
inter-block controlled-not: that is, 127 controlled-nots at the
physical qubit level between ions in separate traps (via the cavities
and fibres). 
Therefore the time per correction of the whole computer is of
order $0.2$ to $0.6$ s. We need about 8 such corrections per Toffoli gate
in the logical algorithm \cite{99:SteaneB}, so the whole algorithm would
take two to eight weeks to run.

There are several methods which could be adopted to reduce this run time
for the motional gates in the ion trap processor. The trap could be made
tighter (increased $\omega_z$): the ions at the centre of the string
would then be too close to be addressed individually, therefore one would use
every other ion in this part of the string. 
There must be more ions in the trap, which off-sets the speed-up, 
but overall a factor 2 speed-up is readily obtained, and a higher
factor if the spacer ions are of a lighter element. 
More demanding techniques offer further
speed-up, for example by using more than one vibrational mode
simultaneously, thus partially parallelizing the gates within a trap.
Furthermore, fault-tolerant methods are based on synthesis of
highly entangled states \cite{97:SteaneA} and make much use
of controlled-multiple-not operations. Such operations
can be generated in an ion trap in a time which
scales as $N^{1/2}$ rather than $N$ as assumed above
\cite{00:Sackett,00:Sorensen,98:Steinbach}.

Fabrication of the electrodes for hundreds of ion traps is
fairly straightforward, using microfabrication methods or otherwise, as
are the low-noise r.f. electronics to provide the trapping fields.
For a detailed analysis of experimental issues for processing within
each trap we refer the reader to \cite{98:WinelandA,98:WinelandB}
and references therein.
Whereas we have not given a thorough treatment of such issues here,
we believe the parameters for trap tightness, optical addressing,
and heating rates which we have assumed are reasonable.
Less well understood is the phenomenon of
charge build-up on the optical cavity mirrors, which will
influence the operation of an ion trap, and
techniques to prevent this may be essential. To build the mirrors for
the optical cavities would be a major undertaking, but a possible
one. One problem would be slow degradation of the mirrors during
assembly or during loading of the traps. To place
the mirrors and traps accurately together, and include associated
optics such as optical fibres, would be taxing but feasible.
The construction and operation of such a quantum computer would have more
in common with the construction of a detector in high energy physics
than with the manufacture of a classical computer chip; it is a
lengthy, expensive and intricate process, but
one whose results might merit the investment of resources, if serious
uses for a 100-qubit computer can be found. The most important point
is that it is conceivable. 

In conclusion, ion trap methods currently offer the only way to achieve
multi-particle entangled states in a controllable way. They offer the
prospect in the fairly near term of achieving various fundamental
principles of quantum information physics, such as
entanglement-enhanced communication and repeatable quantum error
correction. It is clear that the controlled coupling of a trapped ion or
neutral atom to a single-photon field in a high-finesse cavity
also merits investigation for both quantum communication and quantum
computing experiments.

The rough sketch of a quantum computer design which we have given
has two significant features: first, it is based on simple physical
systems whose behaviour, including decoherence mechanisms,
is well understood, and secondly it assumes
only currently accessible levels of technology: all
the components could be built now. The major unknowns are the
detailed experimental issues which may arise when small
optical cavities are combined with ion traps, whether
the approximate error correction analysis gives a fair estimate
of the noise tolerance, and whether
suitable low-level error correction protocols, together with
high-stability laser systems and magnetic and electric field
noise suppression, can provide the $10^{-4}$ gate precision
and $10^{-6}$ memory precision required throughout the computer.
Further experiments, and more thorough feasibility studies, are
certainly called for in this area.

This work was supported by EPSRC, by Christ Church, Oxford
and by the European Community network ``QUBITS''.
We thank D. Stacey for helpful comments on the manuscript.
 

\newpage
\onecolumn

\end{document}